\documentclass[prl,twocolumn,superscriptaddress,amsmath,amssymb]{revtex4-2}

\usepackage{graphicx}
\usepackage{dcolumn}
\usepackage{bm}
\usepackage{titlesec}
\usepackage{hyperref}
\usepackage{appendix}

\usepackage{xcolor}

\def\be{\begin{equation}}
\def\ee{\end{equation}}
\newcommand{\bei}{\begin{itemize}}
\newcommand{\eei}{\end{itemize}}

\newcommand{\hb}{\hbar}
\newcommand{\tb}{\bar{t}}
\newcommand{\csh}{{\rm csch}}

\begin{document}

\title{High-temperature limit penalizing high-frequency quantum fluctuations}

\author{Graeme Pleasance}
\email{gpleasance1@gmail.com}
\affiliation{Department of Physics, University of Stellenbosch, Stellenbosch, 7600, South Africa}
\affiliation{National Institute for Theoretical and Computational Sciences (NITheCS), Stellenbosch, South Africa}
\author{Erik Aurell}
\email{eaurell@kth.se}
\affiliation{KTH -- Royal Institute of Technology, AlbaNova University Center, SE-106 91 Stockholm, Sweden}
\author{Francesco Petruccione}
\email{petruccione@sun.ac.za}
\affiliation{Department of Physics, University of Stellenbosch, Stellenbosch, 7600, South Africa}
\affiliation{National Institute for Theoretical and Computational Sciences (NITheCS), Stellenbosch, South Africa}
\affiliation{School of Data Science and Computational Thinking, University of Stellenbosch, Stellenbosch 7600, South Africa}
\date{\today}

\begin{abstract}
We revisit the Caldeira-Leggett model of quantum Brownian motion with Ohmic spectral density, and derive an additional contribution to the decoherence kernel in a new high-temperature limit at arbitrarily large cut-off frequency. This contribution reveals a novel mechanism for the classicalization of high-frequency quantum fluctuations. We further demonstrate that it leads to a Markovian master equation that is in guaranteed Lindblad form, and argue that this master equation describes the correct Markovian limit of quantum Brownian motion. Our approach considers in detail the behavior of the decoherence kernel at both the initial and final times of the process on the time scale of the bath memory.
\end{abstract}

\maketitle

{\it Introduction.---}The time evolution of a quantum system interacting with an Ohmic bath of harmonic oscillators is one of the most widely studied models of open quantum systems theory \cite{BP2002}. It is applicable in numerous physical situations where the damping becomes friction-like in the semiclassical limit \cite{Weiss2011}, such as for photon and phonon baths \cite{Wilner2015}, as well as superconducting qubits coupled to resistive elements in quantum circuits \cite{Leppakangas2018}. Caldeira and Leggett famously used this model to investigate the classical limit of quantum Brownian motion~\cite{CaldeiraLeggettA} and macroscopic tunnelling~\cite{Caldeira1981, CaldeiraLeggettB}. It also played a central role in later investigations of the spin-boson problem \cite{SpinBoson} and decoherence \cite{Unruh1989,Zurek1993,Paz1993}, as well as recently being used to study gravitational noise effects in the arm lengths of the LIGO interferometer arising from interactions with a quantized bath of gravitons \cite{ParikhWilczekZahariade2020}.

Under a path integral treatment \cite{Kleinert2009}, the influence of the bath splits into dissipation, where the memory is limited by the spectral cut-off $\Omega$ of the interactions, and decoherence, which is also limited by the inverse thermal time $k_B T/\hb$. For phenomenological models and interactions with phonon baths, $\Omega$ is a real physical parameter while for interactions with photons it is a regularizer that can eventually be taken to infinity in renormalization \cite{Breuer2001,Baym2009}. In either case, dissipation is memoryless. In the Caldeira-Leggett (CL) master equation \cite{CaldeiraLeggettA} this dissipation accounts for a renormalization of the system Hamiltonian and two Lindblad terms, both of which are formed of ordered pairs of position and momentum operators $(\hat{X},\hat{P})$ and $(\hat{P},\hat{X})$.

Decoherence is different, in that it has memory. Provided that the thermal time $\hb/k_B T$ is much longer than $\Omega^{-1}$, the cut-off $\Omega$ plays a negligible role, and can be directly taken to infinity.
As described in the pioneering paper by Feynman and Vernon \cite{FeynmanVernon}, decoherence is indistinguishable from averaging over a classical Gaussian field $\xi$ with the same force-force correlations as the memory kernel of the bath. The combined effects of dissipation and decoherence as outlined above are the ingredients of the stochastic Liouville simulations method of Stockburger and Mak \cite{StockburgerMak1999} (see also \cite{Stockburger1998,Stockburger2002}). The decoherence memory kernel $k_r$ is therefore of great importance both in fundamental understanding of the influence of harmonic baths, and in practical simulation tools applied to quantum technologies. Similarly, its role in characterizing thermal fluctuations is particularly relevant to semiclassical descriptions of quantum Brownian motion at high temperatures.

In the standard high-temperature limit \cite{CaldeiraLeggettA}, the kernel $k_r$ has no memory and produces Lindblad terms derived from ordered pairings of the position operator $\hat{X}$ with itself, i.e. $(\hat{X},\hat{X})$.
The classical random force $\xi$ is then delta-correlated in time. In this Letter, we show that if correctly carried out, there is an additional $(\hat{P},\hat{P})$ term in the master equation that serves to both (i) damp high-frequency quantum fluctuations, and (ii) maintain the positivity of the reduced system density operator for parameters consistent with the Born-Markov approximations, similar to what has been identified in previous works \cite{BP2002,Diosi1993,Diosi1993a,Halliwell1995,Gao1997,Ford1999,Vacchini2000,Petruccione2005}. However, our derivation considers a new high-temperature limit which applies to an arbitrarily large (ultraviolet) cut-off $\Omega$ at fixed $T$, and in fact leads to a master equation that is completely independent of $\Omega$. Our result is therefore valid over a broader temperature range than for analogous master equations derived in \cite{BP2002, Diosi1993,Diosi1993a}, and offers new insights into the study of quantum Brownian motion at medium to low temperatures in the Markovian limit. 

{\it Feynman-Vernon theory.---}We focus on the CL model \cite{CaldeiraLeggettA} describing a quantum particle of mass $M$, momentum $\hat{P}$, and position $\hat{X}$, moving in a potential $V(\hat{X})$. The particle interacts with a thermal bath of  harmonic oscillators at (inverse) temperature $T$ ($\beta = 1/k_BT$). As such, the total Hamiltonian reads 
\be\label{eq:H}
\hat{H} = \hat{H}_S + \hat{H}_B + \hat{H}_I + \hat{H}_C,
\ee
where $\hat{H}_S=\hat{P}^2/2M+V(\hat{X})$ and $\hat{H}_B =\sum_k\big(\hat{p}_k^2/2m_k + \frac{1}{2}m_k\omega^2_k\hat{x}_k^2\big)$ are the system and bath Hamiltonians, respectively. The interaction part is taken to have a conventional bilinear form $\hat{H}_I=-\hat{X}
\otimes\sum_kc_k\hat{x}_k$, with $c_k$ denoting the system-bath coupling strength. The final part represents an additional counter term $\hat{H}_C=\hat{X}^2\sum_k(c^2_k/2m_k\omega^2_k)$ that ensures the translational invariance of the Hamiltonian \eqref{eq:H}. 

After integrating out the bath degrees of freedom, the time evolution of the reduced system density operator $\langle X_f|\hat{\rho}_S(t)|X'_f\rangle$ is captured by the propagator $\mathcal{J}(t_f,t_i)$ connecting the initial $X_i$ and final $X_f$ positions of the particle at times $t_i$ and $t_f$. The propagator is expressed as a double path integral over the forward path $X$ from $X_i$ to $X_f$ and backward path $X'$ from $X'_i$ to $X'_f$ \cite{Weiss2011},  
\be
\label{eq:FV-Weiss}
    \mathcal{J}(t_f,t_i) = \int\mathcal{D}X\mathcal{D}X'e^{(i/\hb)(S_S[X] - S_S[X'])}\mathcal{F}[X,X'].
\ee
Here, $S_S[X]=\int^{t_f}_{t_i}dt\,[\frac{1}{2}M\dot{X}^2-V(X)]$ is the free particle action along the path $X(t)$, and $\mathcal{F}[X,X']$ is the Feynman-Vernon (FV) influence functional. Since the bath obeys Gaussian statistics, the influence functional can be written in the exponential form $\mathcal{F}[X,X']= e^{-\frac{i}{\hb}S_i[X,X']-\frac{1}{\hb}S_r[X,X']}$. The quantities $S_r[X,X']$ and $S_i[X,X']$, which couple forward and backward paths, define the real (decoherence) and imaginary (dissipation) parts of the FV action. In terms of the quasiclassical path $\bar{X}(t)=X(t)+X'(t)$ and fluctuation path $\Delta X(t)=X(t)-X'(t)$ \cite{Weiss2011}, they are written as
\begin{align}
S_r[\Delta X] &= \int^{t_f}_{t_i}dt\int^{t}_{t_i}ds\,k_r(t-s)\Delta X(t)\Delta X(s),\label{eq:noise_action} \\
S_i[\bar{X},\Delta X] &= \int^{t_f}_{t_i}dt\int^{t}_{t_i}ds\,k_i(t-s)\Delta{X}(t)\bar{X}(s)\nonumber\\
&\qquad + \mu\int^{t_f}_{t_i}dt\,\Delta X(t)\bar{X}(t),
\end{align}
with $\mu = \sum_kc^2_k/(2m_k\omega^2_k)$ the bath reorganization energy ($\mu\sim\Omega$). The kernels $k_r$ and $k_i$ denote the real and imaginary parts of the bath correlation function $k(\tau)=\sum_k(c^2_k/2m_k\omega_k)[\coth(\hb\omega_k/2k_BT)\cos{(\omega_k\tau)} - i\sin{(\omega_k\tau)}]$. The real part---the FV kernel \cite{FeynmanVernon}---governs the decay of quantum fluctuations along the off-diagonal paths $\Delta X(t)$. In the FV influence functional, it plays the role of a force-force correlation function for a Gaussian stochastic force with memory time $\tau_B$ [see Eq. \eqref{eq:FV_trick} below]. For a bath with Ohmic spectral density $J(\omega)=2\eta\omega\theta(\Omega-\omega)$, where $J(\omega) = \frac{\pi}{2}\sum_k(c^2_k/m_k\omega_k)\delta(\omega-\omega_k)$, the FV kernel reads
\be\label{eq:FV}
    k_r(\tau) = \frac{2\eta}{\pi}\int^{\Omega}_0d\omega\,\omega\coth{\left(\frac{\hb\omega}{2k_BT}\right)}\cos{\omega\tau}.
\ee
It is well known that the white-noise limit of $k_r$ is realized at high-temperature \cite{CaldeiraLeggettA}. In particular, let $\tau_B=\hb/k_BT$. By approximating $\coth{(\hb\beta\omega/2)}\approx 2/\hb\beta\omega$ in the limit $\tau_B\rightarrow 0$, one obtains the the singular kernel $k^{(0)}_r(\tau)=(4\eta/\tau_B)\delta(\tau)$, provided that the particle evolves on time scales much longer than $\Omega^{-1}$. 

\begin{figure}[t!]
\includegraphics[width=0.95\columnwidth]{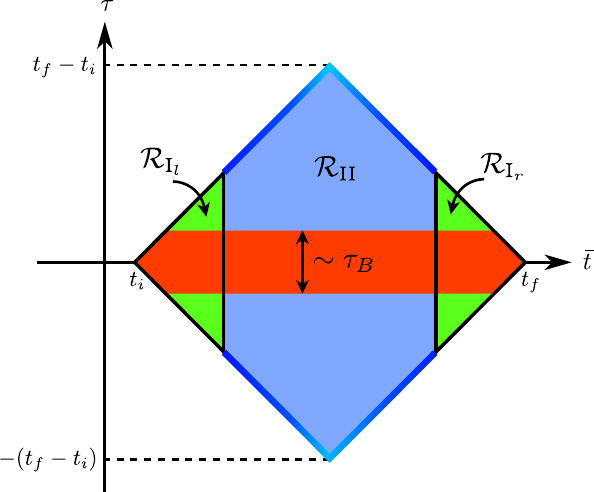}
\caption{Schematic depicting the full domain of integration and sub-domains $\mathcal{R}_{\rm I}=\mathcal{R}_{{\rm I}_l}\cup\mathcal{R}_{{\rm I}_r}$ and $\mathcal{R}_{\rm II}$ of Eq. \eqref{eq:decoh_action}. Since the kernel is essentially zero outside the red-shaded area where $\tau \gg \tau_B$, the $\tau$-integration limits of the blue-shaded area may be extended to $\pm\infty$ assuming $|t_f-t_i|\gg\tau_B$. The green-shaded areas indicate the support of the end contributions where the integration limits cannot be taken to infinity. A more detailed figure illustrating the boundaries of all three sub-domains is provided in \cite{SuppMat}.}
\label{fig1}
\end{figure}

{\it Feynman-Vernon kernel and decoherence action.---}We proceed to now derive a new high-temperature limit for which the FV kernel provides a nonsingular contribution to the decoherence action. As we will show, this leads to an additional diffusion-type term in the CL master equation that serves to damp high-frequency quantum fluctuations. Our approach throughout this Letter is to treat the FV kernel as a tempered distribution, $k_r\in\mathcal{S}'(\mathbb{R})$ \footnote{Here we denote by $\mathcal{S}'(\mathbb{R})$ the space of tempered distributions over $\mathbb{R}$---the continuous dual of the space of Schwartz functions $\mathcal{S}(\mathbb{R})$.}, where the corresponding test functions belong to $\mathcal{S}(\mathbb{R})$ and are {\it a priori} assumed to exhibit negligible variation on time scales shorter than $\Omega^{-1}$. For this purpose, it is convenient to employ the following representation of the FV kernel \cite{SuppMat}
\be\label{eq:kern_func}
	k_r(\tau) =  \frac{2\eta}{\pi}\left[-\frac{\pi^2}{\tau_B^2}\csh^2(\pi \tau/\tau_B) + \Omega^2\,O\left(\frac{1}{(\Omega \tau)^3}\right)\right].
\ee
From this expression it is clear that $k_r$ exhibits two regimes of behavior:
\bei
	\item[(i)] Vacuum regime $\Omega^{-1}\ll \tau\ll \tau_B$: using the asymptotic expansion $\csh^2 x = \frac{1}{x^2}-\frac{1}{3} + O(x^2)$, Eq. (\ref{eq:kern_func}) may be approximated as
	\be\label{eq:kern_short_time}
		k_r(\tau) \approx -\frac{2\eta}{\pi}\frac{1}{\tau^2}, 
	\ee
	implying that the short-time behavior of $k_r(\tau)$ is dominated by the vacuum contribution
    $k^{\rm vac}_r(\tau) = \frac{2\eta}{\pi}\int^{\Omega}_0d\omega\,\omega\cos{\omega \tau}$. In particular, 
    \be
    k^{\rm vac}_r(\tau)=\frac{2\eta}{\pi}\left[-\frac{1}{\tau^2} + \Omega^2O\left(\frac{1}{(\Omega\tau)^3}\right)\right].
    \ee
	\item[(ii)] Exponential regime $\tau\gg \tau_B \gg \Omega^{-1}$: using that $\csh^2 x \sim e^{-2x}$ for $x\gg1$, we get
	\be\label{eq:kern_exp_long_time}
		k_r(\tau) \approx  -\frac{2\eta}{\pi}\frac{1}{\tau^2_B}e^{-2 |\tau|/\tau_B}.
	\ee
\eei
These relations indicate the kernel \eqref{eq:FV} to converge in the scaling limit $\Omega\rightarrow\infty$ for $\tau>0$ and to vanish beyond times $|\tau|\gtrsim\tau_B$ determined by the inverse bath temperature.

With the goal of evaluating the decoherence functional \eqref{eq:noise_action}, we first choose to rewrite the double integral in terms of the new variables $\tb=\frac{1}{2}(t+s)$ and $\tau=t-s$: 
\be\label{eq:decoh_action}
    S_r[\Delta X] = \frac{1}{2}\iint_{\mathcal{R}}d\tb\,d\tau\,k_r(\tau)\Delta X(\tb+\tau/2)\Delta X(\tb-\tau/2).
\ee
The integration region $\mathcal{R}$ in the $(\tb,\tau)$-plane is depicted in Fig. \ref{fig1}. It comprises three subregions $\mathcal{R}_{{\rm I}_l}$ (left), $\mathcal{R}_{{\rm I}_r}$ (right), and $\mathcal{R}_{\rm II}$, where the first two subregions represent the end term contributions to the decoherence action. To simplify the analysis, we choose here to evaluate $S_r[\Delta X]$ without the end terms, which can be shown to play a negligible role at high temperature. A more detailed analysis that explicitly treats these end contributions is provided in Sec. III of the Supplemental Material \cite{SuppMat}. Henceforth, the decoherence action shall be approximated by
\begin{equation}\label{eq:decoh_func}
    S_r[x] \approx \frac{\eta}{\pi}\frac{1}{\tau_B}\int^{t_f}_{t_i}d\tb\int^{\infty}_{-\infty}ds\,F_{\Omega\tau_B}(s)x_{\tb}(s), 
\end{equation}
where $x_{\tb}(s)=\Delta X(\tb+\tau_Bs/2)\Delta X(\tb-\tau_Bs/2)$, and $F_{\Lambda}(s) = \int^{\Lambda}_0ds\,x\coth\frac{x}{2}\cos{sx}$. 

By repeatedly integrating Eq. \eqref{eq:decoh_func} 
by parts, which is permissible since $F_{\Lambda}$ is tempered, we are able to separate the inner integral above into two sets of terms: those depending on the leading contribution to the FV kernel given in Eq. \eqref{eq:kern_func}, and boundary terms depending on the higher order contributions. In particular, the latter are shown to vanish in the scaling limit $\Omega\rightarrow \infty$ (see Sec. II of the Supplemental Material \cite{SuppMat}). By then retaining only the leading terms, and defining the memory kernel $k^{(m)}_r(\tau) = (\pi^2/\tau^2_B)\csh^2(\pi \tau/\tau_B)$, 
the decoherence action can be expressed as the sum of the following two functionals:
\begin{align}\label{eq:action}
	S_r[\Delta X] &:= S^{(0)}_r[\Delta X] + S^{(m)}_r[\Delta X] \nonumber\\
    &=\frac{2\eta}{\pi}\int^{t_f}_{t_i}dt\bigg[\frac{\pi}{\tau_B} \Delta X^2(t)\nonumber\\
    &\hspace{-37pt} + \int^{\infty}_0d\tau\,k^{(m)}_r(\tau)\big(\Delta X^2(t) - \Delta X(t + \tau/2)\Delta X(t - \tau/2)\big)\bigg].
\end{align}
The first functional $S^{(0)}_r[\Delta X]$ is equivalent to the standard high-temperature result obtained for the decoherence action using the memoryless kernel $k^{(0)}_r$. More interestingly, the second functional $S^{(m)}_r[\Delta X]$ depending on $k^{(m)}_r$ is valid beyond the white-noise limit, and therefore represents a contribution to the decoherence action accounting for short-time memory effects. This is exemplified by the fact that $S^{(m)}_r[\Delta X]$ vanishes when $\tau_B\rightarrow 0$, which is directly seen from expanding the path differences inside the inner brackets to lowest order in $\tau$. Hence, by including higher order terms within the expansion of the fluctuations paths, we are able to systematically derive corrections to the white-noise action $S^{(0)}_r[\Delta X]$.

\begin{table*}
\begin{ruledtabular}
\begin{tabular}{cccccc}
 Master equation&Parameters&Lindblad form&$D_{XX}$
&$D_{PP}$&$D_{XP}$\\ \hline
 {\bf This work}&$\hb\gamma\ll k_BT\ll\hb\Omega$&Yes&$\frac{\gamma}{6Mk_BT}$&$\frac{2\gamma Mk_BT}{\hb^2}$ \rule{0pt}{2.4ex}&0 \\
 Di\'{o}si \cite{Diosi1993,Diosi1993a}&$\hb\gamma\ll\hb\Omega\leq k_BT$&Yes&$\frac{\gamma}{6Mk_BT}$&$\frac{2\gamma Mk_BT}{\hb^2}$&$\frac{\Omega\gamma}{6\pi k_BT}$\\
 Breuer-Petruccione \cite{BP2002}&$\hb\gamma\ll\hb\Omega\leq k_BT$&Yes&$\frac{\gamma}{8Mk_BT}$&$\frac{2\gamma Mk_BT}{\hb^2}$&$-\frac{\gamma k_BT}{\hb^2\Omega}$\\
Caldeira-Leggett \cite{CaldeiraLeggettA}&$\hb\gamma\ll\hb\Omega\leq k_BT$&No&0&$\frac{2\gamma M k_BT}{\hb^2}$&0
\end{tabular}
\end{ruledtabular}
\caption{\label{tab1} Comparison between different Markovian master equations derived in the high-temperature limit with corresponding diffusion coefficients \cite{Homa2019}. In \cite{Diosi1993,Diosi1993a}, the CL master equation is rederived in Lindblad form for certain parameter limits. Note also that the cross diffusion term $D_{PX}$ derived in \cite{BP2002} (Eq. (3.409) therein) is treated as negligible compared to $D_{PP}$ when the system evolution is slow compared to $\Omega^{-1}$.}
\end{table*}

{\it New high-temperature limit.---}We proceed to expand $x_{t}(s)$ to next leading order 
in $\tau_Bs$ (or $\tau$), 
\be\label{eq:x_expand}
    x_{t}(s) = x^{(0)}_{t} + (\tau_Bs)^2x^{(1)}_{t} + ...,
\ee
with $x^{(0)}_{t}=\Delta X^2_{t}$, such that Eq. \eqref{eq:action} becomes
\begin{equation}\label{eq:action_exp}
    S_r[\Delta X] = S^{(0)}_r[\Delta X] + S^{(1)}_r[\Delta X] + ...,
\end{equation}
where
\begin{align}\label{eq:decoh_action_1}
    S^{(1)}_r[\Delta X] &:= \frac{\eta\tau_B}{6}\int^{t_f}_{t_i}dt\,\Delta X(t)\left(-\frac{\partial^2}{\partial t^2}\right)\Delta X(t) \\
    &+ ({\rm boundary \,\, terms}). \nonumber
\end{align}
Note that here we have used $\int^{\infty}_0d\tau/\tau_B\,\tau^2k^{(m)}_r(\tau) = \pi/6$. Equation \eqref{eq:decoh_action_1} is the first result of this Letter and provides the leading high-temperature correction to the white-noise action. We may elucidate its role within the functional \eqref{eq:decoh_action} by rewriting $\exp(-S_r[\Delta X]/\hb)$ as the average over an auxiliary Gaussian field $\xi(t)$, where $\mathbb{E}[\xi(t)]=0$ and $\mathbb{E}[\xi(t)\xi(s)]=k^{\rm eff}_r(t-s)$. For this, let us introduce the differential operator $K(t)=\frac{1}{\sigma^2}\big(-\frac{\partial^2}{\partial t^2} + \lambda^2\big)$ with $\lambda=2\sqrt{3}/\tau_B$ and $\sigma=\sqrt{3/\eta\tau_B}$, such that up to the leading order correction of Eq. \eqref{eq:action_exp} we have $\exp(-S_r[\Delta X]/\hb) = \exp(-\frac{1}{2\hb}\int^{t_f}_{t_i}dt\,\Delta X(t)K[\Delta X](t))$. This exponential can then be recast using a Hubbard-Stratonovich transformation as \cite{StockburgerMak1999,Xu2023,Xu2025}
\begin{align}\label{eq:FV_trick}
	&\exp{(-S_r[\Delta X]/\hb)}=({\rm trivial})\nonumber\\
    &\quad\times\int \mathcal{D}\xi\,P[\xi]\exp\left(-\frac{i}{\hb}\int^{t_f}_{t_i}dt\,\xi(t)\Delta X(t)\right),
\end{align}
where $P[\xi] = N^{-1}\exp(-\frac{1}{2\hb}\int^{t_f}_{t_i}dt\,\xi(t)K^{-1}[\xi](t))$ is a Gaussian probability functional. The trivial prefactor in this expression stems from the boundary terms of Eq. \eqref{eq:decoh_action_1}. Additionally, $K^{-1}[\xi] = \int^{t_f}_{t_i}ds\,G(t-s)\xi(s)$, where $G(t-s)$ is the (symmetric) Green's function of the operator $K(t)$.

Since this kernel must be positive definite and satisfy $\frac{1}{\sigma^2}\big(-\frac{\partial^2}{\partial t^2} + \lambda^2\big)G(t-s)=\delta(t-s)$, it is straightforward to deduce that $k^{\rm eff}_r(t-s)=K(t-s)\delta(t-s)$. Notably, this force-force correlation function is equivalent to the FV kernel expanded to its next leading term in the high-temperature limit, i.e. $k_r(\tau)\approx \frac{4\eta}{\tau_B}\delta(\tau) - \frac{\eta\tau_B}{3}\delta''(\tau)$. This implies the correction term $S^{(1)}_r[\Delta X]$ to be valid at the same level of approximation made by Di\'{o}si \cite{Diosi1993,Diosi1993a}, where an analogous correction was derived by treating the FV kernel to $O(\hb\beta)$. The correction there was shown to lead to an additional diffusion term in the CL master equation rendering it in Lindblad form \cite{Diosi1993,Diosi1993a}. In parallel, we demonstrate that our result also gives rise to a Lindblad form master equation but with Dekker coefficients \cite{Dekker1977,Dekker1981} that are independent of the cut-off $\Omega$.

{\it Derivation of the Lindblad master equation from the Feynman-Vernon kernel.---}For a Liouvillian $\mathcal{L}$ that generates a completely positive and uniformly continuous dynamical semigroup, its most general representation is in Lindblad form \cite{Gorini1976, Lindblad1976}
\begin{align}\label{eq:me} 
&\frac{\partial\hat{\rho}_S(t)}{\partial t} = \mathcal{L}\hat{\rho}_S(t) \nonumber\\ 
&\hspace{-.11cm}= -\frac{i}{\hb}[\hat{H}_{\rm eff},\hat{\rho}_S]+ \sum_{i,j} a_{ij}
\Big(\hat{L}_i\hat{\rho}_S\hat{L}_j^{\dagger}
-\frac{1}{2}\{\hat{\rho}_S,
\hat{L}_j^{\dagger} \hat{L}_i\}\Big),
\end{align}
where $\hat{L}_i$ are Lindblad operators, and $\hat{H}_{\rm eff}$ is a Hamiltonian \footnote{The ``eff'' subscript on $H_{\rm eff}$ indicates that under a microscopic derivation of Eq. \eqref{eq:me}, the Hamiltonian contribution can contain additional Lamb shifts and/or frequency renormalization terms not contained in $\hat{H}_S$.}. The positivity of the solution obtained from this equation is guaranteed given the matrix $\boldsymbol{a}$ is necessarily positive semidefinite, $\boldsymbol{a}\succeq 0$. In \cite{BP2002, CaldeiraLeggettA}, the CL master equation derived under the Born-Markov approximations has the same form as Eq. \eqref{eq:me} with Lindblad operators $\hat{L}_1=\hat{X}$, $\hat{L}_2=\hat{P}$, and Kossakowski matrix 
\be\label{eq:a0}
\boldsymbol{a}_0=
\begin{pmatrix}
  4\gamma Mk_BT/\hb^2 & -i\gamma/\hb \\
  i\gamma/\hb & 0
\end{pmatrix}, \quad \gamma=\eta/M
\ee
whose spectrum always possesses a negative eigenvalue (${\rm det}\,\boldsymbol{a}_0<0$). The reason for this discrepancy is a missing $(\hat{P},\hat{P})$ term which emerges from the next leading contribution in the high-temperature expansion of the FV kernel \cite{BP2002,Diosi1993,Diosi1993a,Gao1997,Ford1999,Vacchini2000}. In our case, rewriting the FV influence functional in superoperator representation \cite{Aurell2020} gives the Liouvillian 
\begin{align}\label{eq:liouville}
	\mathcal{L}\hat{\rho}_S &= -\frac{i}{\hb}[\hat{H}_S,\hat{\rho}_S] - i\frac{\gamma}{\hb}[\hat{X},\{\hat{P},\hat{\rho}_S\}] - D_{PP}[\hat{X},[\hat{X},\hat{\rho}_S]] \nonumber\\
					    &\quad - D_{XX}[\hat{P},[\hat{P},\hat{\rho}_S]] - 2D_{XP}[\hat{X},[\hat{P},\hat{\rho}_S]],
\end{align}
where
\be\label{eq:dc}
	D_{XX} = \frac{\gamma}{6Mk_BT}, \quad D_{PP} = \frac{2\gamma Mk_BT}{\hb^2}, \quad D_{XP} = 0,
\ee
are phenomenological Dekker coefficients \cite{Dekker1977,Dekker1981}. We remark that \eqref{eq:liouville} holds for any form of potential $V(\hat{X})$ contained in $\hat{H}_S$, which solely influences the particle's free evolution and not the diffusion timescales \eqref{eq:dc} \cite{AlickiLendi2007}. By setting $\hat{L}_1=\hat{X}$, $\hat{L}_2=\hat{P}$, it is straightforward to show that the Kossakowski matrix obtained from the Liouvillian \eqref{eq:liouville} satisfies the positivity constraint through ${\rm det}\,\boldsymbol{a} = 4D_{XX}D_{PP} - \frac{\gamma^2}{\hb^2} \geq0$. As such, the position diffusion coefficient $D_{XX}$ is sufficiently large to always render the master equation in Lindblad form. We further remark that since the additional $(\hat{P},\hat{P})$ term damps high-frequency quantum fluctuations, at sufficiently high frequency the quantum system interacting with the bath behaves classically---a new open quantum system avenue to \textit{classicalization} \cite{Dvali2011}.

Table \ref{tab1} summarizes the key differences between the Markovian master equation derived in this work [Eqs. \eqref{eq:liouville}-\eqref{eq:dc}] and those derived in Refs. \cite{Diosi1993,Diosi1993a,CaldeiraLeggettA,BP2002}. %

{\it Connection to exact master equations.---}Finally, we note that Eq. \eqref{eq:me} with $\hat{L}_1=\hat{X}$, $\hat{L}_2=\hat{P}$, and $\hat{H}_{\rm eff}=\hat{P}^2/2M + \frac{1}{2}M\omega^2_0\hat{X}^2$, has an inherently different structure to the exact Hu-Paz-Zhang (HPZ) master equation \cite{Hu1992} describing non-Markovian quantum Brownian motion (see also \cite{Strunz1996,DiosiStrunz1997,DiosiGisinStrunz1998,Yu2004,StrunzYu2004,DiosiFerialdi2014,Ferialdi2016} for similar exact results). This is due to the $(\hat{P},\hat{P})$ term which has no direct counterpart in the HPZ master equation. In Refs. \cite{Ferialdi2016,Ferialdi2017}, it was argued that this discrepancy stems from dissipation being an inherently non-Markovian feature of the CL model, and that dissipation in the Markovian limit can only be accounted for 
by introducing a momentum-mediated coupling to the bath \cite{Diosi1995,Halliwell1995,Gao1997,Vacchini2000}. In particular, a phenomenological approach including both a position and momentum coupling to an external noise was shown to unravel the time-local master equation \cite{Ferialdi2016}
\begin{align}\label{eq:tl_liouvillian}
    \mathcal{L}_t\hat{\rho}_S &= -i[\hat{H}_{\rm eff}(t),\hat{\rho}_S] + \Gamma(t)[\hat{X},[\hat{X},\hat{\rho}_S]] + \Theta(t)[\hat{X},[\hat{P},\hat{\rho}_S]] \nonumber\\
    &+ \Xi(t)[\hat{X}^2,\hat{\rho}_S] + \Upsilon(t)[\hat{X},\{\hat{P},\hat{\rho}_S\}] + \gamma(t)[\hat{P},[\hat{P},\hat{\rho}_S]]
\end{align}
where a $(\hat{P},\hat{P})$ term does appear. Since Eq. \eqref{eq:tl_liouvillian} reduces to \eqref{eq:liouville} in the white-noise limit, it was concluded that a $\hat{P}$-coupling is required to correctly describe quantum Brownian motion in the Markovian regime.

Alternatively, we argue that Markovian quantum Brownian motion is best described within the new high-temperature limit introduced in this Letter. This is based on the fact that for the other master equations listed in Table \ref{tab1}, taking the Markovian limit $\Omega\rightarrow\infty$ also requires $T\rightarrow\infty$, which amounts to retaining only the leading contribution to the FV kernel, i.e. $k_r\sim\delta(\tau)$. This type of kernel leads to an unphysical description where the complete positivity of the dynamics is lost. In our approach, however, the next-leading contribution to $k_r$ can be retained while also enforcing a truly memoryless bath $\Omega\rightarrow\infty$, suggesting that this provides the correct implementation of the Markovian limit. Ultimately, Eq. \eqref{eq:liouville} with a harmonic potential $V(\hat{X}) = \frac{1}{2}M\omega^2_0\hat{X}^2$ {\it cannot} be recovered as limiting case of the HPZ master equation \cite{BP2002}, thereby distinguishing our result from \cite{Hu1992,StrunzYu2004,Yu2004}.

It is also worth noting that Eq. \eqref{eq:liouville} admits an unraveling into the stochastic Schr\"{o}dinger equation \cite{StrunzYu2004}
\begin{equation}
    \dot{\psi}_t = \Big(-\frac{i}{\hbar}\hat{H}_S +\sum_i\hat{L}_i\phi_i(t) - \frac{1}{2}\sum_{i,j}a_{ij}\hat{L}^{\dagger}_j\hat{L}_i\Big)\psi_t,
\end{equation}
with $\phi_i(t)$ zero-mean Gaussian white noise, and $\mathbb{E}[\phi_i(t)\phi^*_j(s)]=a_{ij}\delta(t-s), \quad \mathbb{E}[\phi_i(t)\phi_j(s)] = 0$. This equation contains a momentum-noise coupling term $\hat{P}\phi(t)$ akin to that phenomenologically introduced in \cite{Ferialdi2016} As such, the inclusion of the next-leading term of the FV kernel $\sim\delta''(\tau)$ is sufficient to introduce the effective $\hat{P}$-coupling required to maintain complete positivity in line with Refs. \cite{Halliwell1995, Gao1997,Ford1999,Vacchini2000,Petruccione2005,Diosi1995}.

{\it Conclusions.---}In this Letter, we have derived a new high-temperature limit of the CL model describing a quantum particle interacting with an Ohmic bath of harmonic oscillators. The reduced system dynamics in this limit has shown to be captured by a Markovian master equation in Lindblad form, in contrast to the CL master equation used in the conventional white-noise treatment of quantum Brownian motion \cite{CaldeiraLeggettA}, among other similar works \cite{Diosi1993,Diosi1993a,Halliwell1995,Gao1997,Ford1999,Vacchini2000,Petruccione2005}. 
Furthermore, the master equation \eqref{eq:liouville} applies to any form of potential $V(\hat{X})$, and does not require an additional momentum coupling to the bath to describe completely positive quantum Brownian motion. Since our result also applies to baths with an ultraviolet cutoff $\Omega$, it is likely to be applicable outside the temperature regime stipulated in Refs. \cite{BP2002,CaldeiraLeggettA,Diosi1993} (c.f. Table \ref{tab1}), provided that the temperature is still large enough for the Born-Markov approximation $\hb\gamma\ll k_BT$ to be satisfied. A more formal comparison between these master equations could be explored for a harmonic oscillator system by benchmarking their predictions against the exact HPZ master equation \cite{Hu1992}. Furthermore, our result may be systematically extended to lower temperature regimes by including higher order terms within the expansion of the fluctuation paths \eqref{eq:x_expand}, which is a task left to future work.

\bibliographystyle{apsrev4-2} 

\bibliography{References}

\appendix
\widetext
\pagebreak

\begin{center}
\textbf{\large Supplemental Material for ``A high-temperature limit penalizing high-frequency quantum fluctuations''}\\[1em]
Graeme Pleasance, Erik Aurell, and Francesco Petruccione 
\end{center}
\vspace{1em}

\titleformat{\section}{\bfseries\centering\uppercase}{\thesection.}{1em}{}
\titleformat{\subsection}{\bfseries}{\thesubsection}{1em}{}

\renewcommand{\thesection}{\Roman{section}}
\renewcommand{\thesubsection}{\Alph{subsection}}

\section{I. Derivation of the Feynman-Vernon kernel}

In this section we outline the derivation of the FV kernel given in Eq. \eqref{eq:kern_func} of the main text. Our approach is based partly on the derivation presented in Ref. \cite{BP2002} to evaluate the decoherence functional for a relativistic particle interacting with an Ohmic bath---see Eq. (12.172) therein---which for the sake of completeness, is reproduced here in full. 

We start by defining the function 
\be\label{eq:F}
    F_{\Lambda}(s) = \int^{\Lambda}_0dx\,x\coth\frac{x}{2}\cos{sx},
\ee
which is related to the FV kernel according to ($\tau_B = \hb\beta$)
\begin{align}\label{eq:fv_kern}
	k_r(\tau) &= \frac{2\eta}{\pi}\Big(\frac{1}{\hb\beta}\Big)^2\int^{\Omega\hb\beta}_0dx\,x\coth\frac{x}{2}\cos\frac{x\tau}{\hb\beta} \nonumber\\
		 &= \frac{2\eta}{\pi}\frac{1}{\tau_B^2} F_{\Omega\tau_B}(\tau/\tau_B).
\end{align}
We further introduce the first and second primitives of Eq. \eqref{eq:F}, 
\begin{align}
	G_{\Lambda}(s) &= \int^{\Lambda}_0dx\,\coth\frac{x}{2}\sin sx, \\
	H_{\Lambda}(s) &= \int^{\Lambda}_0dx\,\coth\frac{x}{2}\left(\frac{1-\cos sx}{x}\right),
\end{align}
with 
\be\label{eq:primatives}
	\frac{d^2}{ds^2}H_{\Lambda}(s) = \frac{d}{ds}G_{\Lambda}(s) = F_{\Lambda}(s),
\ee
and divide $H_{\Lambda}(s)$ into a vacuum and thermal part $H^{\rm vac}_{\Lambda}(s)$ and $H^{\rm th}_{\Lambda}(s)$ \cite{BP2002}:
\begin{align}
	H_{\Lambda}(s) &:= H^{\rm vac}_{\Lambda}(s) + H^{\rm th}_{\Lambda}(s) \nonumber\\
    &= \int^{\Lambda}_0dx\frac{1-\cos sx}{x} + \int^{\Lambda}_0dx\left(\coth\frac{x}{2}-1\right)\frac{1-\cos sx}{x}.
\end{align}
Now, for the vacuum contribution it follows that
\begin{align}\label{eq:H_vac}
	H^{\rm vac}_{\Lambda}(s) &= \int^{\Lambda s}_0dx\frac{1-\cos x}{x} \nonumber\\
					  &= {\rm ln}\,g\Lambda s + \int^{\infty}_{\Lambda s}dx\frac{\cos x}{x} = {\rm ln}\,g\Lambda s + O\left(\frac{1}{\Lambda s}\right),
\end{align}
where the second equality has been obtained through the identity $\int^{\infty}_{\Lambda s}dx\frac{\cos x}{x} = -\ln g\Lambda s + \int^{\Lambda s}_0dx\frac{1-\cos x}{x}$, and $\ln g\approx 0.577$ is the Euler-Mascheroni constant. Upon differentiating Eq. (\ref{eq:H_vac}), we also obtain the vacuum terms
\begin{align}
	G^{\rm vac}_{\Lambda}(s) &= \frac{1}{s} + \Lambda \,O\left(\frac{1}{(\Lambda s)^2}\right) \\
	F^{\rm vac}_{\Lambda}(s) &=  -\frac{1}{s^2} + \Lambda^2\,O\left(\frac{1}{(\Lambda s)^3}\right).
\end{align}
For the thermal component of $G_{\Lambda}(s)$,
\be
	G^{\rm th}_{\Lambda}(s) = \int^{\Lambda}_0dx\left(\coth\frac{x}{2} -1\right)\sin sx,
\ee
we may extend the upper limit of the integration to infinity, since $(\coth\frac{x}{2}-1)$ falls off exponentially for $x>1$. It therefore follows from the identity
\[
	\coth\frac{x}{2} -1 = 2\sum^{\infty}_{n=0}e^{-(n+1)x}
\]
that
\begin{align}\label{eq:G_therm}
	G^{\rm th}_{\Lambda}(s) &\approx 2\,{\rm Im}\int^{\infty}_0dx\sum^{\infty}_{n=0}e^{isx - (n+1)x} \nonumber\\
    &= \pi\coth \pi s - \frac{1}{s},
\end{align}
where the second equality is obtained from the Matsubara decomposition
\[
	\coth\frac{x}{2} = 2\left(\frac{1}{x} + 2\sum^{\infty}_{n=1}\frac{x}{x^2+(2\pi n)^2}\right).
\]
Next we may combine $G^{\rm vac}_{\Lambda}(s)$ with Eq. \eqref{eq:G_therm}, yielding
\be\label{eq:G}
	G_{\Lambda}(s) = \pi\coth\pi s + \Lambda O\left(\frac{1}{(\Lambda s)^2}\right).
\ee
The thermal part of $F_{\Lambda}(s)$ may also be obtained by means of differentiating Eq. (\ref{eq:G}),
\be
	F^{\rm th}_{\Lambda}(s) = -\frac{\pi^2}{\sinh^2\pi s} + \frac{1}{s^2},
\ee
so that by combining both vacuum and thermal parts of $F_{\Lambda}(s)$, we get
\be\label{eq:FL}
	F_{\Lambda}(s) = -\pi^2\csh^2\pi s + \Lambda^2O\left(\frac{1}{(\Lambda s)^3}\right).
\ee
Returning to Eq. \eqref{eq:fv_kern}, the FV kernel is then given as
\be
	k_r(\tau) =  \frac{2\eta}{\pi}\left[-\frac{\pi^2}{\tau_B^2}\csh^2(\pi \tau/\tau_B) + \Omega^2\,O\left(\frac{1}{(\Omega \tau)^3}\right)\right].
\ee

\section{II. Evaluation of the decoherence functional}\label{sec:functional}

In this section we derive the explicit form of the decoherence functional given in Eq. \eqref{eq:action} of the main text which excludes the end term contributions. Our starting point is to introduce the functional $T_F:\mathcal{S}(\mathbb{R})\rightarrow\mathbb{R}$,
\be\label{eq:functional}
T_F[x_{\tb}] = \int^{\infty}_{-\infty}ds\,F_{\Lambda}(s)x_{\tb}(s),
\ee
which acts on the space of test functions $\mathcal{S}(\mathbb{R})$ that includes the set of fluctuation paths $x_{\tb}(s)=\Delta X(\tb+\tau_Bs/2)\Delta X(\tb - \tau_Bs/2)$. As stated in the main text, these paths are assumed to also exhibit negligible variation on time scales shorter than $\Omega^{-1}$. We proceed to partition the integral \eqref{eq:functional} into three disjoint subintervals set by the parameter $q$ (see Fig. \ref{fig2}),
\be\label{eq:functional_supp}
T_F[x] = \int^{q}_{-q}ds\,F_{\Lambda}(s)x(s) + \int^{\infty}_qds\,F_{\Lambda}(s)x(s) + \int^{-q}_{-\infty}ds\,F_{\Lambda}(s)x(s),
\ee
where the subscript on $x_{\tb}$ has been dropped for ease of notation (the $\tb$-dependence of $x_{\tb}$ will be implicit hereon unless stated). By expanding $x(s)=x(0)+O(s)$, the first integral may be written as
\begin{align}\label{eq:func_q}
    \int^q_{-q}ds\,F_{\Lambda}(s)x(s) &= \int^{\Lambda}_0dy\,y\coth\frac{y}{2}\bigg(\int^{q}_{-q}ds\,x(s)\cos{sy}\bigg) \nonumber\\
    &= x(0)\int^{\Lambda}_0dy\,y\coth\frac{y}{2}\bigg(\int^q_{-q}ds\,\cos{sy}\bigg) + ... \nonumber\\
    &= 2G_{\Lambda}(q)x(0) + ... \, .
\end{align}
Now assuming $\Lambda q\gg1$, we can use the leading term of Eqs. \eqref{eq:G} and \eqref{eq:func_q} to infer that the distribution $F_{\Lambda}(s)$ behaves as a Dirac delta on the interval $[-q,q]$ for sufficiently small $q$:
\[
    x\mapsto \int^q_{-q}ds\,F_{\Lambda}(s)x(s)= 2\pi\coth{(\pi q)}x(0),\qquad q\gg\Lambda^{-1}.
\]
As such, we may also use Eq. \eqref{eq:FL} to write 
\be
    T_F[x] = 2\pi\int^q_{-q}ds\,x(s)\delta(s)\coth{(\pi q)} - \pi^2\bigg[\int^{\infty}_qds\,x(s)\csh^2{(\pi s)} + \int^{-q}_{-\infty}ds\,x(s)\csh^2{(\pi s)}\bigg], 
\ee
where integrating by parts terms in the square brackets yields
\begin{align}
    T_F[x] &= 2\pi\coth{(\pi q)}\int^q_{-q}ds\,x(s)[\delta(s) - \delta(s+q) - \delta(s-q)] \nonumber\\
    &\qquad - \pi\int^{\infty}_qds\,\coth{(\pi s)}[x'(s) - x'(-s)].
\end{align}
Applying integration by parts twice again to the second line of the above then leads to
\be\label{eq:functional_ibp}
    T_F[x] = 2\pi\,\coth{(\pi q)}[x(0) - x(q)] - 2\pi\int^{\infty}_qds\,\coth{(\pi s)}\,x'(s), 
\ee
having used that $x(s)$ is an even function. To evaluate the remaining integral the derivative $x'(s)$ can be written as 
\be\label{eq:deriv}
    x'(s) = \frac{d}{ds}[x(s) - x(0)].
\ee
This form of $x'(s)$ is needed else we end up with a diverging term in the result (since $\coth{(\pi s)}$ is not integrable at the origin). Accordingly, by integrating the second term of Eq. \eqref{eq:functional_ibp} by parts, we get 
\be
    T_F[x] =  2\pi\bigg(x(0) - \pi\int^{\infty}_qds\,\csh^2{(\pi s)}[x(s) - x(0)]\bigg).
\ee
Now taking the limit $q\rightarrow 0$, which is justified by simultaneously taking $\Lambda\rightarrow \infty$ ($\Omega\rightarrow\infty$) to ensure $\Lambda q \gg 1$, (this can be controlled by e.g. fixing $q\sim \Lambda^{-\alpha}$, $0<\alpha<1$), one obtains 
\be\label{eq:T_lim}
    \lim_{\Lambda\rightarrow\infty}\lim_{q\rightarrow 0}T_F[x_{\tb}] =  2\pi\bigg(x_{\tb}(0) - \pi\int^{\infty}_0ds\,\csh^2(\pi s)[x_{\tb}(s) - x_{\tb}(0)]\bigg),
\ee
having reintroduced the $\tb$ subscript on $x$. It should be noted that the integral converges in this limit since the derivative \eqref{eq:deriv} vanishes at the origin. Finally using 
\be
    S_r[x] = \frac{\eta}{\pi}\frac{1}{\tau_B}\int^{t_f}_{t_i}d\tb\,T_F[x_{\tb}]
\ee 
and inserting \eqref{eq:T_lim} yields Eq. \eqref{eq:action} of the main text.

\section{III. Evaluation of the end terms of the decoherence functional}\label{sec:end_terms}

\begin{figure}[h!]
\includegraphics[width=.57\textwidth]{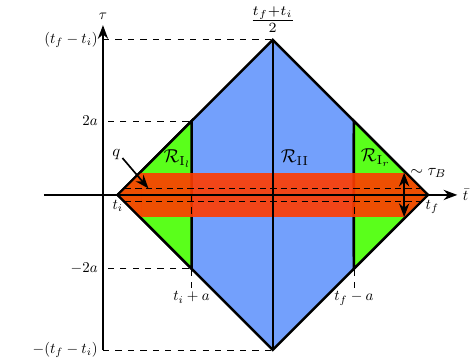}
\caption{A more detailed schematic of the integration region presented in Fig. \ref{fig1} of the main text. Each of the subregions $\mathcal{R}_{\rm I}$ and $\mathcal{R}_{\rm II}$ are colored green and blue, respectively. The red region indicates the part of the $(\tb,\tau)$-plane where the FV kernel $k_r(\tau)$ has nonvanishing support. In subregion $\mathcal{R}_{\rm II}$, the integration domain in $\tau$ can extended to infinity and treated as per the main text. Section III concerns the treatment of the corner regions $\mathcal{R}_{{\rm I}_l}$ and $\mathcal{R}_{{\rm I}_r}$ parameterized by $a$. It is assumed that the test functions $x_{\tb}$ have negligible variation on time scales less than $a$.}
\label{fig2} 
\end{figure}

Here we seek to properly treat the end term contributions to the decoherence action given in Eq. \eqref{eq:decoh_action} of the main text. In terms of the integration variables $\tb$ and $\tau$, the functional reads
\be\label{eq:dec_func_end_terms}
	S_r[\Delta X] = \frac{1}{2}\iint_{\mathcal{R}_{\rm{I}} + \mathcal{R}_{\rm{II}}}d\tb\,d\tau\,k_r(\tau)\Delta X(\tb+\tau/2)\Delta X(\tb-\tau/2),
\ee
where $\mathcal{R}_{\rm{I}/\rm{II}}$ define the green and blue colored regions shown in Fig. \ref{fig2}: 
\begin{align*}
	&\mathcal{R}_{\rm I} = \mathcal{R}_{{\rm I}_l} \cup \mathcal{R}_{{\rm I}_r}\\
	&\qquad \mathcal{R}_{{\rm I}_l} = \{(\tb,\tau) :  t_i\leq \tb\leq t_i+a, \, -2(\tb-t_i)\leq\tau\leq 2(\tb-t_i)\} \\
	&\qquad \mathcal{R}_{{\rm I}_r} = \{(\tb,\tau) :  t_f-a\leq \tb\leq t_f, \, -2(t_f-\tb)\leq\tau\leq 2(t_f-\tb)\} \\
	{\rm and} \\
	&\mathcal{R}_{\rm II} = \mathcal{R}_{{\rm II}_l} \cup \mathcal{R}_{{\rm II}_r}\\
	&\qquad \mathcal{R}_{\rm {II}_l} = \{(\tb,\tau) : t_i+a\leq \tb\leq \frac{1}{2}(t_f+t_i), \, -2(\tb-t_i)\leq\tau\leq 2(\tb-t_i)\}  \\
	&\qquad \mathcal{R}_{\rm {II}_r} = \{(\tb,\tau) : \frac{1}{2}(t_f+t_i)\leq \tb\leq t_f-a, \, -2(t_f-\tb)\leq\tau\leq 2(t_f-\tb)\}.
\end{align*}	
The parameter $a$ defines the boundary of the region $\mathcal{R}_{\rm II}$ in relation to the end points $t_i$ and $t_f$. If $a$ is fixed such that $a\gtrsim\tau_B$, the $\tau$-integration limits in $\mathcal{R}_{\rm II}$ can be extended to the full real line since we are not picking up any contributions that are significant compared to the red region. The parameter $q$ in this region can also be made arbitrarily small as per Sec. II. Hence, the corresponding decoherence functional takes the same form as Eq. \eqref{eq:action} of the main text with modified $\tb$-limits,
\begin{align}
	S^{\rm II}_r[\Delta X] &= \frac{1}{2}\int^{t_f-a}_{t_i+a}d\tb\int^{\infty}_{-\infty}d\tau\,k_r(\tau)\Delta X(\tb+\tau/2)\Delta X(\tb-\tau/2) \nonumber\\
		    &= \frac{2\eta}{\pi}\int^{t_f-a}_{t_i+a}d\tb\left[\frac{\pi}{\tau_B} \Delta X^2(\tb) + \int^{\infty}_0d\tau\,k^{(m)}_r(\tau)(\Delta X^2(\tb) - \Delta X(\tb+\tau/2)\Delta X(\tb-\tau/2))\right],
\end{align}
which in the high-temperature limit becomes
\be\label{eq:dec_func_II_htl}
	S^{\rm II}_r[\Delta X] \approx \frac{2\eta}{\tau_B}\int^{t_f-a}_{t_i+a}d\tb\,\Delta X(\tb)^2 + \frac{\eta\tau_B}{12}\int^{t_f-a}_{t_i+a}d\tb\,\Big(\Delta X'(\tb)^2 - \Delta X(\tb)\Delta X''(\tb)\Big).
\ee
Now considering the subregion $\mathcal{R}_{\rm I}$, the decoherence action within each corner segment $\mathcal{R}_{{\rm I}_l}$ and $\mathcal{R}_{{\rm I}_r}$ is given by 
\begin{subequations}
\begin{align}
    S^{{\rm I}_l}_r[\Delta X] &= \frac{1}{2}\int^{t_i+a}_{t_i}d\tb\int^{2(\tb-t_i)}_{-2(\tb-t_i)}d\tau\,k_r(\tau)\Delta X(\tb+\tau/2)\Delta X(\tb-\tau/2), \label{eq:decoh_action_l} \\
    S^{{\rm I}_r}_r[\Delta X] &= \frac{1}{2}\int^{t_f}_{t_f-a}d\tb\int^{2(t_f-\tb)}_{-2(t_f-\tb)}d\tau\,k_r(\tau)\Delta X(\tb+\tau/2)\Delta X(\tb-\tau/2). \label{eq:decoh_action_r}
\end{align}
\end{subequations}
We proceed by isolating the contributions to each of the $\tb$-integrals from around the end points $t_i$ and $t_f$:
\begin{subequations}
\begin{align}
    S^{{\rm I}_l}_r[x] &= \bar{S}^{{\rm I}_l}_r[x] + R^{{\rm I}_l}_r[x], \\
    S^{{\rm I}_r}_r[x] &= \bar{S}^{{\rm I}_r}_r[x] + R^{{\rm I}_r}_r[x],
\end{align}
\end{subequations}
where 
\begin{subequations}
\begin{align}
    R^{{\rm I}_l}_r[x] &:= \frac{\eta}{\pi\tau_B}\int^{t^+_i}_{t_i}d\tb\int^{\frac{2(\tb - t_i)}{\tau_B}}_{\frac{-2(\tb - t_i)}{\tau_B}}ds\,F_{\Lambda}(s)x_{\tb}(s), \label{eq:R_left}\\ 
    R^{{\rm I}_r}_r[x] &:= \frac{\eta}{\pi\tau_B}\int^{t_f}_{t^-_f}d\tb\int^{\frac{2(t_f - \tb)}{\tau_B}}_{\frac{-2(t_f - \tb)}{\tau_B}}ds\,F_{\Lambda}(s)x_{\tb}(s),\label{eq:R_right}
\end{align}
\end{subequations}
define the remainder terms. The remainder for the left corner region can be estimated from
\begin{align}
    \left|\int^{t^+_i}_{t_i} d\tb\int^{\frac{2(\tb - t_i)}{\tau_B}}_{-\frac{2(\tb - t_i)}{\tau_B}} ds\,F_{\Lambda}(s)x_{\tb}(s)\right| &\leq \int^{t^+_i}_{t_i} d\tb\int^{\frac{2(\tb - t_i)}{\tau_B}}_{-\frac{2(\tb - t_i)}{\tau_B}} ds\,|F_{\Lambda}(s)x_{\tb}(s)| \nonumber\\
    &\leq M_{\Lambda}\int^{t^+_i}_{t_i} d\tb\int^{\frac{2(\tb - t_i)}{\tau_B}}_{-\frac{2(\tb - t_i)}{\tau_B}}ds \nonumber\\
    &= 2M_{\Lambda}\frac{\delta t^2}{\tau_B},
\end{align}
with $M_{\Lambda} := \sup_{\tb,s\in\mathcal{R}_{{\rm I}_l}}|F_{\Lambda}(s)x_{\tb}(s)|$ and $\delta t := t^+_i-t_i$. This implies 
\be
    |R^{{\rm I}_l}_r| \leq \frac{2\eta M_{\Lambda}}{\pi}\frac{\delta t^2}{\tau^2_B},
\ee
such that $R^{{\rm I}_l}_r[x]$ can be made small as long as $\delta t \ll \sqrt{\pi\tau^2_B/2\eta M_{\Lambda}}$. A similar analysis also applies to $R^{{\rm I}_r}_r[x]$.

Focusing now on the action functionals $\bar{S}^{{\rm I}_{l/r}}_r[\Delta X]$, the inner integrals of Eqs. \eqref{eq:decoh_action_l} and \eqref{eq:decoh_action_r} can be evaluated following the same procedure outlined in Sec. II. In doing so, we get 
\begin{subequations}
\begin{align}
	\bar{S}^{{\rm I}_l}_r[\Delta X] &= \frac{2\eta}{\tau_B}\int^{t_i+a}_{t^+_i}d\tb\,\coth\left(\frac{2\pi(\tb-t_i)}{\tau_B}\right)\Delta X(\tb)^2 \nonumber\\
		      &\qquad + \frac{2\eta}{\pi}\int^{t_i+a}_{t^+_i}d\tb \int^{2(\tb-t_i)}_qd\tau\,k^{(m)}_r(\tau)\big(\Delta X^2(\tb) - \Delta X(\tb + \tau/2)\Delta X(\tb - \tau/2)\big), \label{eq:S_I_left}\\
	\bar{S}^{{\rm I}_r}_r[\Delta X] &= \frac{2\eta}{\tau_B}\int^{t^-_f}_{t_f-a}d\tb\,\coth\left(\frac{2\pi(t_f-\tb)}{\tau_B}\right)\Delta X(\tb)^2 \nonumber\\
		      &\qquad + \frac{2\eta}{\pi}\int^{t^-_f}_{t_f-a}d\tb \int^{2(t_f-\tb)}_qd\tau\,k^{(m)}_r(\tau)\big(\Delta X^2(\tb) - \Delta X(\tb + \tau/2)\Delta X(\tb - \tau/2)\big). \label{eq:S_I_right}
\end{align}
\end{subequations}
Note that unlike Eq. \eqref{eq:T_lim}, the limit $q\rightarrow0$ of the $\tau$-integrals cannot be formally taken since $\lim_{\Lambda\rightarrow\infty}M_{\Lambda}=\infty$, and hence the remainder terms $R^{{\rm I}_{l/r}}_r[x]$ will not be properly bounded. We can, however, assume $q$ to be sufficiently small (or $\Lambda$ sufficiently large) so that the lower limit of these integrals can be approximately taken to zero. 

Let us now consider the high-temperature limit. Expanding the inner terms of Eqs. (\ref{eq:S_I_left}) and (\ref{eq:S_I_right}) to leading order in $\tau$ yields
\begin{subequations}
\begin{align}
	\bar{S}^{{\rm I}_l}_r[\Delta X] &\approx \Big(\dots\Big) + \frac{2\tau_B\eta}{\pi^2}\int^{t_i+a}_{t^+_i}d\tb\, B\left(\frac{2\pi(\tb-t_i)}{\tau_B}\right)\big(\Delta X'(\tb)^2 - \Delta X(\tb)\Delta X''(\tb)\big), \label{eq:S_I_l_htl} \\
	\bar{S}^{{\rm I}_r}_r[\Delta X] &\approx\Big(\dots\Big) + \frac{2\tau_B\eta}{\pi^2}\int^{t^-_f}_{t_f-a}d\tb\,B\left(\frac{2\pi(t_f-\tb)}{\tau_B}\right)\big(\Delta X'(\tb)^2 - \Delta X(\tb)\Delta X''(\tb)\big), \label{eq:S_I_r_htl}
\end{align}
\end{subequations}
where $B(t)$ is defined as the antiderivative of the kernel $k^{(m)}_r(\tau)$,
\be\label{eq:B}
	B\Big(\frac{\pi t}{\tau_B}\Big) := \frac{\pi}{\tau_B}\int^t_0d\tau\,k^{(m)}_r(\tau) = \int^{\frac{\pi t}{\tau_B}}_0dx\,x^2\csh^2x .
\ee
An analytic result for this integral is found to be
\be\label{eq:B_scaled}
        B(t) = t\big[t - t\coth{t} + 2\ln(1-e^{-2t})\big] - {\rm Li}_2(e^{-2t}) + \frac{\pi^2}{6},
\ee
where ${\rm Li}_s(t) = \sum^{\infty}_{k=1}\frac{t^k}{k^s}$ is the polylogarithm function. 

\begin{figure}[t!]
\centering
\includegraphics[width=.5\textwidth]{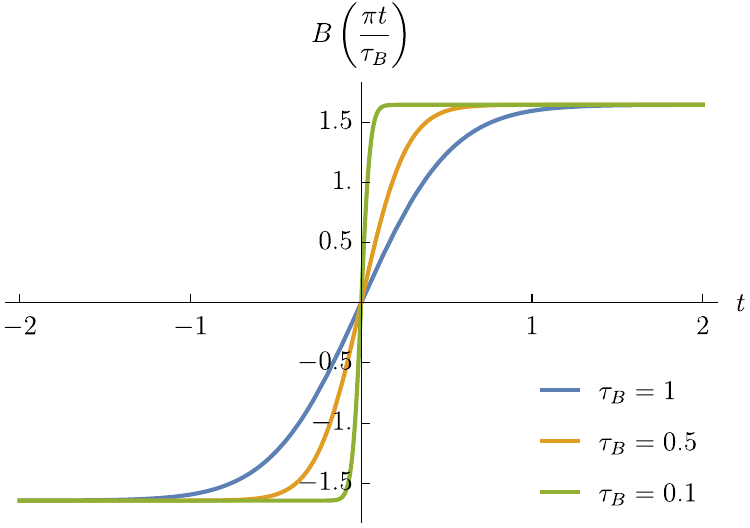}
\caption{\label{fig3} Plot of the integral (\ref{eq:B}) with varying upper limit set by $t$. As $t\rightarrow\infty$ it asymptotically reaches the value $\frac{\pi^2}{6}$; see Eq. (\ref{eq:B_scaled}).}
\end{figure}

In Fig. \ref{fig3}, the integral $B(\pi t/\tau_B)$ is shown for several values of $\tau_B$. The inverse time $\tau^{-1}_B$ determines the rate at which the asymptotic value of the integral is reached. Therefore, when $\tau_B$ is sufficiently small, the inner integral of Eq. (\ref{eq:S_I_left}) will be constant for almost all $\tb$-values except for those close to the boundary $t_i$. The same also applies to the inner integral of Eq. (\ref{eq:S_I_right}) when $\tb$ approaches the upper limit $t_f$.

\begin{figure}[t!]
\centering
\includegraphics[width=.6\textwidth]{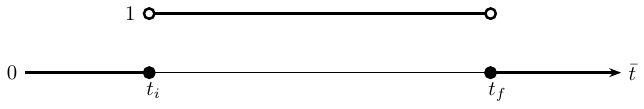}
\caption{\label{fig4} Integration measure $d\mu(\tb)$ acting on the interval $[t_i, t_f]$ in the high-temperature limit.}
\end{figure}

We may formalize this behavior as follows. Let $\mu(\tb)$ define a measure in the piecewise sense
\begin{align}
	d\mu(\tb) = 
	\begin{cases}
		d\tb \qquad\qquad  & \tb\in(t_i+m, t_f-m\tau_B), \\ 
		\frac{6}{\pi^2}B\left(\frac{\tb-t_i}{\tau_B}\right)d\tb \qquad\qquad & \tb\in[t_i, t_i+m\tau_B],  \\ 
		\frac{6}{\pi^2}B\left(\frac{t_f-\tb}{\tau_B}\right)d\tb \qquad\qquad  & \tb\in[t_f-m\tau_B, t_f],    
	\end{cases}
\end{align}
with $m\in\mathbb{R}^+$ chosen such that $B(m\tau_B/\pi)\sim \frac{\pi^2}{6}$. Then as $\tau_B\rightarrow0$, the regions for which $B(\pi t/\tau_B)$ deviates from $\frac{\pi^2}{6}$ will be confined to the end points of the interval $[t_i,t_f]$. Since these points have zero measure, this implies
\be\label{eq:int_measure}
	\lim_{\tau_B\rightarrow0}d\mu(\tb) = 
	\begin{cases}
		d\tb \qquad &\tb\in(t_i,t_f), \\
		0, \qquad &\tb\in\{t_i,t_f\},
	\end{cases}
\ee
where the limiting behavior of $\mu(\tb)$ is illustrated in Fig. \ref{fig4}.

Because Eqs. (\ref{eq:S_I_l_htl}) and (\ref{eq:S_I_r_htl}) are already expanded in the limit of small $\tau_B$, we may write
\begin{align}\label{eq:dec_func_result}
	S_r[\Delta X] &= S^{\rm I}_r[\Delta X] + S^{\rm II}_r[\Delta X] \nonumber\\
	      &\approx \frac{2\eta}{\tau_B}\bigg[\int^{t_f-a}_{t_i+a}d\tb\,\Delta X(\tb)^2  \nonumber \\
	      &\qquad + \int^{t_i+a}_{t^+_i}d\tb\,\coth\left(\frac{2\pi(\tb-t_i)}{\tau_B}\right)\Delta X(\tb)^2 + \int^{t^-_f}_{t_f-a}d\tb\,\coth\left(\frac{2\pi(t_f-\tb)}{\tau_B}\right)\Delta X(\tb)^2\bigg] \nonumber\\
	      &\qquad + \frac{\eta\tau_B}{12}\int^{t_f}_{t_i}d\mu(\tb)\big(\Delta X'(\tb)^2 - \Delta X(\tb)X''(\tb)\big) + R^{{\rm I}_l}_r[\Delta X] + R^{{\rm I}_r}_r[\Delta X].
\end{align}
By further noting in the same limit that 
\begin{align*}
    &\lim_{\tau_B\rightarrow0}\coth\left(\frac{2\pi(\tb-t_i)}{\tau_B}\right) = 1 \qquad \tb\in(t_i,t_i+a), \\
    &\lim_{\tau_B\rightarrow0}\coth\left(\frac{2\pi(t_f-\tb)}{\tau_B}\right) = 1 \qquad \tb\in(t_f-a,t_f), 
\end{align*}
the first and second lines of Eq. \eqref{eq:dec_func_result} may also be written in terms of the measure $d\mu(\tb)$: 
\be
    S_r[\Delta X] \approx \int^{t_f}_{t_i}d\mu(\tb)\bigg[\frac{2\eta}{\tau_B}\Delta X^2(\tb) + \frac{\eta\tau_B}{12}\big(\Delta X'(\tb)^2 - \Delta X(\tb)\Delta X''(\tb)\big)\bigg] + R^{{\rm I}_l}_r[\Delta X] + R^{{\rm I}_r}_r[\Delta X].
\ee
Hence, having explicitly treated the end terms of the decoherence functional $S_r[\Delta X]$, we recover an analogous result to Eqs. \eqref{eq:action}-\eqref{eq:decoh_action_1} of the main text provided the remainder terms can be made negligibly small.

\end{document}